\documentclass[]{acmart}

\usepackage{tabularx}
\usepackage{caption}
\usepackage{subcaption}
\usepackage{array}
\usepackage{todonotes} 
\usepackage{soul}
\newcolumntype{C}[1]{>{\centering\arraybackslash}p{#1}}

\AtBeginDocument{%
  \providecommand\BibTeX{{%
    \normalfont B\kern-0.5em{\scshape i\kern-0.25em b}\kern-0.8em\TeX}}}

\begin{document}

\title{Sparks: Inspiration for Science Writing using Language Models}

\author{Katy Ilonka Gero}
\email{katy@cs.columbia.edu}
\affiliation{%
  \institution{Columbia University}
  \city{New York}
  \state{New York}
  \country{USA}
}

\author{Vivian Liu}
\email{vivian@cs.columbia.edu}
\affiliation{%
  \institution{Columbia University}
  \city{New York}
  \state{New York}
  \country{USA}
}

\author{Lydia B. Chilton}
\email{chilton@cs.columbia.edu}
\affiliation{%
  \institution{Columbia University}
  \city{New York}
  \state{New York}
  \country{USA}
}
\renewcommand{\shortauthors}{Gero, Liu, and Chilton.}

\begin{abstract}
Large-scale language models are rapidly improving, performing well on a wide variety of tasks with little to no customization. In this work we investigate how language models can support science writing, a challenging writing task that is both open-ended and highly constrained. We present a system for generating “sparks”, sentences related to a scientific concept intended to inspire writers. We find that our sparks are more coherent and diverse than a competitive language model baseline, and approach a human-created gold standard. In a study with 13 PhD students writing on topics of their own selection, we find three main use cases of sparks: aiding with crafting detailed sentences, providing interesting angles to engage readers, and demonstrating common reader perspectives. We also report on the various reasons sparks were considered unhelpful, and discuss how we might improve language models as writing support tools.
\end{abstract}



\begin{teaserfigure}
  \includegraphics[width=\textwidth]{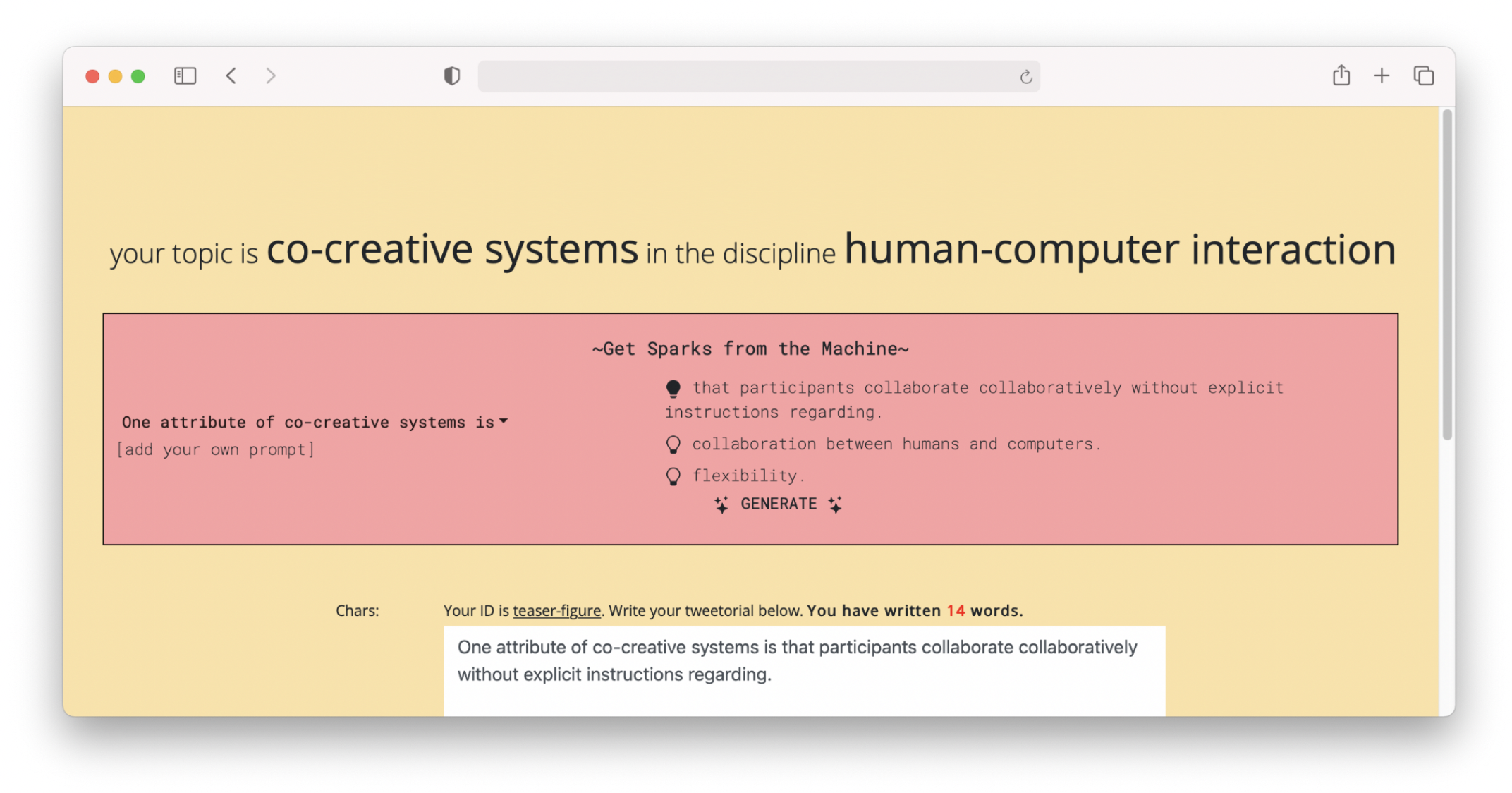}
  \caption{An example screenshot of our system for the topic `co-creative systems' in the discipline `human-computer interaction'. The system has generated three ``sparks": sentences intended to inspire the participant when writing an explanation for their topic. The first spark has been `starred' as inspirational.}
  \Description{adfsdfasdfsdfsad.}
  \label{fig:teaser}
\end{teaserfigure}

\maketitle

\section{Introduction}

New developments in large-scale language models have produced models that are capable of generating coherent, convincing text in a wide variety of domains \cite{vaswani_attention_nodate, brown_language_2020, adiwardana_towards_2020}. Their success has spurred improvements on many tasks, from classification to question answering to summarization \cite{brown_language_2020}, as well as creative writing support \cite{coenen_wordcraft_2021}. 
Language models have the potential to be powerful writing tools that can support writers in real-world, high-impact domains. These new models are task agnostic, making them applicable to many tasks without requiring more training, and we believe such models are the future of AI technologies. 

Despite their successes, language models continue to exhibit known problems, such as generic outputs \cite{holtzman_curious_2020}, lack of diversity in their outputs \cite{ippolito_comparison_2019}, and factually false or contradictory information \cite{lin_truthfulqa_nodate}. There remain many unknowns about how this technology will interface with people in real-world tasks, such as what interactions best serve writers, how language models can best contribute to different writing forms \cite{calderwood_how_2018}, and how to mitigate the bias that language models encode \cite{bender_dangers_2021}. 

To that end, we study how language models can be applied to a real-world, high-impact writing task. In particular, we use a science writing form called “tweetorials” which explain technical concepts on Twitter for a general audience \cite{breu_tweetstorm_2020}. Tweetorials are short explanations of around 500 words which have a low-barrier to entry and are gaining popularity as a science writing medium \cite{soragni_scientists_2019}. 
Working on science writing requires a system to demonstrate proficiency within an area of expertise. This is much more difficult than traditional creative writing tasks, such as stories and poetry, which tend to deal with common objects and relations. Thus, we present the following research question:

\begin{quote}
\textit{RQ: How can outputs from a language model support writers in a creative but constrained writing task?}    
\end{quote}

In this paper, we present a system that aims to inspire domain experts when writing tweetorials on a topic of their expertise. This system provides what we call “sparks” -- sentences intended to spark ideas in the writer. Our system generates sparks using a mid-sized language model (GPT-2 \cite{radford_language_nodate}) and a custom decoding method to encourage specific and diverse outputs. Additionally, we develop a set of 10 prompts based on narratology and expository theory that participants can use when interacting with the model.

We run two evaluations. In the first, we compare the outputs from a custom decoding method to a competitive baseline as well as to a human-created gold standard, reporting on the diversity and coherence of all outputs. In our second study, we have 13 PhD students from five STEM disciplines write tweetorials with our system and report on how they thought about and made use of the sparks.

We make the following contributions:

\begin{itemize}
    \item a system that uses a language model to generate “sparks” related to a scientific concept;
    \item a custom decoding method for generating sparks from a pre-trained language model;
    \item an evaluation
    demonstrating that the sparks are more coherent and diverse than an off-the-shelf system, and approach a human gold standard; and
    \item an exploratory study with 13 PhD students showing three main use cases of sparks.
\end{itemize}

We end by discussing how to best use language models in constrained writing tasks.

\section{Related Work}

\subsection{Natural Language Generation}

A language model is any model that predicts the likelihood of a sequence of words. This can be used to generate text by giving the model a prefix -- either a sequence of words or a special `start of sequence' word -- and having it calculate the likelihood of all words in its vocabulary as the next word. This can be used to select the next word, and thus generate text. Language models are trained on a dataset of text that does not need to be annotated in any way, as the model can simply be trained on what the next word in a sentence will be \cite{stanford-nlp-textbook}.

Language models are getting larger: they are being trained on more text and the models have more parameters \cite{radford_language_nodate,brown_language_2020, adiwardana_towards_2020}. Much recent work has been on how to make the best use of these large language models, which have shown to be much more general purpose than previous ones \cite{keskar_ctrl_2019}, even showing promise in generating code \cite{austin_program_2021}. Additionally, it is useful to be able to take one language model and use it for many tasks, rather than having to train a new model for each task.

It has been shown that a well-selected prefix, or `prompt', can dramatically increase the performance of a language model on a specific task \cite{reynolds_prompt_2021}. A resulting line of research has been automatically creating either natural language prompts or continuous vector prompts, to perform well on tasks \cite{gao_making_2021, li_prefix-tuning_2021}. However, automatically learned prompts have yet to consistently outperform manually crafted prompts \cite{gao_making_2021}. Additionally there have been very few studies on prompt selection for open ended, generative tasks \cite{coenen_wordcraft_2021}.

Despite the successes of language models, problems remain. When considering the generation of text, language models tend to output repetitive and vague responses \cite{holtzman_curious_2020, ippolito_comparison_2019}. Language models also have no model of the truth; they are learning correlations from large amounts of text. Thus they are able to produce text that includes falsehoods and offensive language \cite{bender_dangers_2021}.

\subsection{Generative Writing Support}

Technological writing support has a long history, but has seen an increase in attention as language models have improved. Early work on language models for creative writing focused on activities such as storytelling \cite{roemmele_writing_nodate} and metaphor writing \cite{chakrabarty_mermaid_2021}. While these tools proved helpful for writers, they were narrow in what they could provide. An exploratory study found that generic auto completion from a language model did not provide enough control for novelists \cite{calderwood_how_2018}. More recent work on writing support for creative tasks has varied the ways in which technology can support the writer, for instance by providing description, plot points, or even asking questions, depending on the desires of the writer \cite{coenen_wordcraft_2021, arnold_generative_nodate}.

Writing support for nonfiction writing tasks tends to be much more constrained, for instance as sentence completion \cite{chen_gmail_2019}. A good example is Gmail's smart reply function, which aims to suggest only text that the writer would have written in anyway \cite{kannan_smart_2016}, although it has been shown that even these suggestions can change what people write \cite{arnold_sentiment_nodate}. Work on helping people craft responses to those in mental health crisis focuses on providing writers feedback, and suggested words, rather than complete phrases or sentences \cite{peng_exploring_2020}.

While natural language generation is a large and growing field, few of its technologies are studied in the context of how they will be used by writers. For instance, although there is much work on automatic summarization \cite{zhang_pegasus_2020, he_ctrlsum_2020}, there's less work on how the summaries might be used by people. Our work aims to study how text generated by language models might be used by writers in a science writing task. There's some relation to a natural language generation task like summarization, because we are concerned with real facts, but we take a human centered approach where the language model provides suggestions, rather than a completed output.

\subsection{Science Communication on Social Media}

Science communication helps the public understand scientific contributions -- consider how it has been applied to tackle vaccine misinformation \cite{shelby_story_2013}, the COVID-19 pandemic \cite{yang_covid-19_2020}, and climate change \cite{hart_boomerang_2012}. Traditionally, science communication took place through journals, conferences, articles, and books -- places where peer review was an implicit part of the publication process. However, the rise of digital networks has made science establish a virtual presence through electronic journals and digital records. The ubiquity of social media further presented opportunities for scientists to have direct channels to the public. Now any scientist can conduct science communication online by posting about their work online \cite{soragni_scientists_2019}, engaging in the `Ask' communities on Reddit \cite{gilbert_i_2020} or explaining something on Youtube \cite{welbourne_science_nodate}. Even PhD students or undergraduate researchers have the ability to disseminate their scientific knowledge at any time without depending on a venue or a publication.

This emerging trend, where the scientist can now partake in conversations outside of an implicitly gated, peer-review process, reflects one of the many broad shifts away from traditional science communication. Scholars of science communication have reified this emerging form of communication as ``post-normal science communication'' \cite{bruggemann_post-normal_2020}. Defining characteristics of post-normal science communication include a tolerance for subjectivity, an insertion of the self, the integration of advocacy, and call to actions. Despite these dramatic shifts, the original tenets of science communication such as storytelling, analogies, figures, and citations remain valuable, and storytelling in particular is a driving principle within our system.

\subsection{Expository and Narrative Theory}
In studying how narratives are embedded in text, we turn to a rich body of literature about narratives and knowledge structure in semiotics and discourse theory. These domains inform our search for structures we could use to prompt language models.
We looked at frameworks for both expository and narrative writing, because tweetorials are a hybrid of both. Specifically, we draw from the constructionist theory for narrative text, discourse theory for narrative text, and discourse theory for expository text.

The constructionist framework of narratology states that all reading comprehension is “a search for meaning'' \cite{Graesser1994}. Readers infer as they build a mental model of why certain actions, events, and states are involved in a situation. The constructionist framework has a classification of inferences that we borrow from for many of our prompt templates. Our prompts exemplify a subset of these inference classes such as case structure role assignment, causal antecedent, the presence of superordinate goals, and instantiation of a noun category. 

Concurrently, we examined expository text discourse theory for knowledge structures that would lend well to prompt templates. One framework for expository text introduced a taxonomy of didactic methods (evaluation, explanation, occasion, and expansion) to enumerate the different conversational moves a writer can make to "influence the inference process of the reader" \cite{Tucker1986}. An alternative and popular framework was put forth by Meyer et. al., who enumerated signal phrases that distinguish expository texts, such as ‘specifically’ or ‘attributes of’. We chose to incorporate multiple signal phrases from Meyer’s framework into our prompt templates \cite{meyer_structure_2017}.

\section{Formative Study}

In order to understand how a language model might best support the task of writing a tweetorial, we ran a formative study where participants were first given a technique for coming up with a compelling introduction, before being asked to write the first tweet of a tweetorial on a technical topic they were familiar with. Since the first tweet tends to set up the context and intention of the tweetorial \cite{breu_tweetstorm_2020} we found this to be an effective and efficient way to understand what participants found difficult in the writing process, even when provided with writing strategies.

\subsection{Methodology}

We recruited 10 students in Computer Science.\footnote{6 women / 4 men; 7 undergraduates--no first years / 3 PhD students.}
Participants were required to go through tutorials on how to write an engaging science writing introduction on two example topics -- recursion and virtual private networks -- which included several examples and a step-by-step process for coming up with ideas. These tutorials focused on coming up with an intriguing question for the first tweet, and were developed in consultation with a science journalist.
The process was: 1) brainstorm three concrete situations related to the topic, 2) turn each situation into a question for the reader, 3) select the most engaging question.\footnote{Links to tutorials will be released after anonymous review.} These tutorials were intended to provide the participants with as much “unintelligent” support as possible, such that we could identify where language models may be able to add benefit.

After the tutorials, participants were asked to select a topic from one of six Computer Science topics and write the first tweet for a tweetorial that would explain that topic.\footnote{The topics were: hashing, sorting algorithms, Bayes theorem, HTTP, transistors, and Turning Machines. We selected these topics as ones that a) most computer science students should have learned in a formal setting, and b) could make for an interesting tweetorial.} Participants were asked to think aloud during the writing process. They were not allowed to browse the web. Afterwards, they were asked a series of questions about their writing process in a semi-structured interview.

After all participants had completed the study, the research team reviewed their writing with a science journalist. No formal coding was done, but general areas of success and areas for development were discussed.

\subsection{Results}

\subsubsection{Participants reported that the task required creativity, and that it was difficult to come up with ideas.}

Although we didn’t frame the task as a creative writing task, many participants described the task as difficult because it required creativity to come up with something that would engage the reader. Most participants said they don’t typically do creative writing, so they found the task difficult and outside of their area of comfort. This supported our selection of tweetorials as a writing task, as we want to study a task that is both constrained and creative.

Participants found the tutorials helpful, though for a variety of reasons. Some liked seeing the examples, others appreciated a process to follow, and still others found it comforting to see writing get better with brainstorming and revision. Several commented that the tutorials made the task look easy, but when they began to write about their own topic it was surprisingly difficult to come up with ideas.

Most participants (9 out of 10) said that making the topic interesting to a general audience was the most difficult part of the writing task. When pressed to be more specific, participants mentioned coming up with concrete examples/situations and creating an engaging question as hard tasks. Though this was surely influenced by the process the tutorials introduced, this confirmed that tutorials are not enough to fully support writers in this task.

\subsubsection{Participants struggled to come up with ideas that created suspense.}

When reviewing what the participants had written, all the tweets mimicked the tone of the examples. However, the science journalist had critiques for all of them, and most of the critiques at the core were the same: the tweet lacked suspense. By this he meant, the tweet did not introduce a compelling problem or gap in the reader’s understanding that would make the reader want to read more. Often this was because the example used wasn’t particularly compelling or didn’t reflect a real use case of the topic.
Additionally, participants tended to repeat similar ideas to others who had selected the same topic. For instance, all the people writing about HTTP used either Google or Twitter as their example, suggesting that participants may converge on similar, easy to reach ideas.

Given that participants reported coming up with ideas difficult, it's likely that participants could have done better if given help coming up with more ideas.
Members of the research team also noted that many of the tweets written might be difficult to turn into full length tweetorials. For instance, if the question couldn’t really be answered with an explanation about their chosen topic. For this reason, in future studies we had participants write more than just the first tweet.

\subsection{Design Goals}

Based on our formative study, we developed two design goals for our system:

\subsubsection{Support writers with idea generation.}

Given that language models have no model of truth, we want our system to come up with “sparks”, intended to spark ideas in the writer, rather than having the system provide the ideas themselves. This aligns with prior work on creativity support tools, where users make use of system outputs as initial directions that are then interpreted and diverged from in the users' actual creation \cite{gero_metaphoria_2019}.
Additionally, this also encourages the writer to feel more ownership over their final product, which has shown to be a concern in past work \cite{peng_exploring_2020}.

\subsubsection{Generate outputs that are coherent and diverse.}

In order for writers to make use of outputs, even if they are not always perfectly accurate, they should be coherent -- well-formed and generally reflecting accurate knowledge. Additionally, to support idea generation, outputs should also be diverse, such that writers have a variety of outputs to make use of.

\section{System Design}

\subsection{Generating Sparks}

\subsubsection{Language model selection.}

To generate sparks we use GPT-2, an open source, mid-sized (1.5 billion parameters), transformer language model trained on 40GB of text from the web \cite{radford_language_nodate}. We use the huggingface implementation \cite{wolf_transformers_2020}. While larger open source models are available, e.g. GPT-3 \cite{brown_language_2020} or Megatron-LM \cite{shoeybi_megatron-lm_2020}, we wanted to limit the size of the model we used as larger models are more expensive to run and take more time to generate text. Additionally, there have been many critiques of the super-large language models \cite{bender_dangers_2021}, and thus we wanted to use the smallest language model able to perform well for our use case. Anecdotally, we found that DistilGPT2, a `distilled', smaller version of GPT-2 \cite{sanh_distilbert_2020}, was not able to produce coherent responses to our prompts.
We experimented with fine-tuning GPT-2 on a data set of science writing, but found that this made little difference, especially compared to modifying the decoding method or the prompts. For this reason most of our design effort focused on decoding and prompt engineering.

\subsubsection{Decoding method.}

In addition to selecting a model, we had to design a decoding method -- how to select the next token given the probability distribution the model outputs. There are several common ways of decoding from language models: greedy search, beam search, top-k sampling \cite{holtzman_curious_2020}, and top-n sampling \cite{fan_hierarchical_2018}, to name a few. Different methods have different strengths and weaknesses. Beam search tends to produce high quality results \cite{meister_if_2021} but also tends to produce very similar results for the same prompt. Sampling methods can produce much more varied results, but at the cost of being less coherent. We designed a method that attempts to further increase the coherence of beam search while also increasing its diversity.

\begin{figure}
  \includegraphics[width=.6\textwidth]{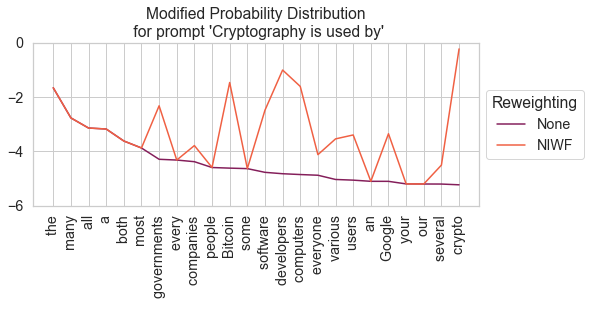}
  \caption{This graph shows how the likelihood of the 20 next most likely words given the prompt "cryptography is used by". The orange line shows the distribution after it has been rewieghted with normalized inverse word frequencies (NIWF). Words like ``governments'', ``Bitcoin'', ``software'', and ``developers'' have an increased probability, while words like ``many'', ``both'', and ``all'' are not modified.}
  \Description{adfsdfasdfsdfsad.}
  \label{fig:logits}
\end{figure}

First, we modify the probability distribution using a normalized inverse word frequency, in order to increase the likelihood of infrequent words. Normalized inverse word frequency is often used in natural language generation to improve the specificity of outputs \cite{ko_domain_nodate, zhang_learning_nodate}, which is one method for increasing the overall quality of results. To our knowledge this is the first work to use normalized inverse word frequency purely as a decoding method as opposed to during training. 
To calculate the word frequencies, we wanted a corpus that doesn't over-represent uncommon science words, like a science writing dataset might, but also reflects modern word usage. For these reasons,
we use a corpus of Vox news articles that includes all articles published before March, 2017.\footnote{\url{https://data.world/elenadata/vox-articles}}
\autoref{fig:logits} shows an example of the probability distribution being modified. In this figure you can see that words like ``governments'', ``Bitcoin'', ``software'', and ``developers'' have an increased weight, while words like ``many'', ``both'', and ``all'', are not modified.

Second, we use only the top 50 highest ranking tokens. This is sometimes called top-k sampling, as only the top $k$ tokens are used \cite{fan_hierarchical_2018}. However, since we're not using a sampling method, the effect of this is to ensure that the modified probability distribution doesn't introduce any incoherencies, for example by dramatically increasing the rank of a token very far down in the original probability distributions. 

Third, we increase the diversity of outputs by forcing the first token of each output to be unique, but attempt to retain coherence generating the rest of the tokens with beam search. While several more sophisticated methods have been proposed to increase diversity while retaining the coherence of beam search (e.g. \cite{vijayakumar_diverse_2018}), in testing we found none were as effective as simply enforcing the first token to be unique.

Finally, in order to keep the sparks succinct and generating quickly, we only generate 10 tokens after the prompt, and cut off the generation as soon as a sentence has been completed.

We implement our decoding method using the huggingface transformers \cite{wolf_transformers_2020}.\footnote{Link to code to be added after anonymous review.}

\subsubsection{Prompt design.}

Designing prompts for language models has become an active area of research, with many automatic methods being proposed \cite{gao_making_2021, li_prefix-tuning_2021}. However, any automatic method requires at least some training data, and  it's yet to be seen that automatically developed prompts can outperform hand-crafted prompts \cite{gao_making_2021}. For these reasons, we hand-craft our prompts.

First we craft a `prefix' prompt to pre-pend to any prompt used by a writer. Prefix prompts have been shown to greatly improve performance by providing the language model with appropriate context \cite{reynolds_prompt_2021}. We found early on in development that simply providing the model with a technical topic was not enough -- also providing a context area was necessary for it to appropriately interpret technical terms. For instance, if you use a prompt like "Natural language generation is used for", the model is likely to talk about linguistic research on languages, rather than computational methods. If instead you use the prompt, "Natural language generation, a topic in computer science, is used by" the results are much more likely to refer to computational language generation. Given this, we pre-pend all prompts with the following:
``\{topic\} is an important topic in \{context area\}''
where \{topic\} and \{context area\} are provided by the writer.

In hand-crafting our prompts, we wanted to make sure our prompts captured a range of relevant angles, so our system could flexibly work with any technical discipline. To do so, we synthesized work from expository and narrative theory into prompts capturing five categories: expository, instantiation, goal, causal, and role. Each category represented an angle that a writer might want to explore.  All prompts can be seen in \autoref{tab:prompts}.

\begin{table}
  \caption{Prompt templates designed for science writing task.}
  \label{tab:prompts}
  \begin{tabular}{ll}
    \toprule
    category & prompt\\
    \midrule
     expository 
    &One attribute of \{topic\} is\\
    &Specifically, \{topic\} has qualities such as\\
    instantiation
    &One application of \{topic\} in the real world is \\
    &\{topic\} occurs in the real world when\\
    goal
    &For instance, people use \{topic\} to \\
    &\{topic\} is used for\\
    causal
    &\{topic\} happen because \\
    &For example, \{topic\} causes\\
    role
    &\{topic\} is used by \\
    &\{topic\} is studied by\\
  \bottomrule
\end{tabular}
\end{table}

We manually developed these prompts according to established frameworks within narrative and expository theory that we referenced in our related work. Our prompts within the categories of instantiation, goal, antecedent, and role were crafted based upon the constructionist framework of inferences, specifically the following categories: case structure role assignment, causal antecedent, the presence of superordinate goals, and the instantiation of a noun category (respectively). Less formally, \textit{Instantiation} prompt templates suggest completions that instantiate where and in what ways topic X may occur in the real world. \textit{Goals} prompt templates suggest completions that represent how topic X is used in the real world. \textit{Causes} prompt templates suggest completions for how topic X might interact in cause and effect chains. \textit{Roles} prompt templates cover entities involved with topic X. 
As tweetorials exhibit both elements of narrative and expository writing, we also borrowed signal phrases from Meyer’s framework for expository text \cite{meyer_structure_2017} -- e.g. “specifically", ``such as", ``attribute” -- and folded them within our prompt templates.

In testing we found that participants often wanted to `follow up' on an output by entering in their own prompt. For this reason, we added the ability for writers to add their own prompts, though this prompt would also be pre-pended with our prefix.\footnote{One intriguing area of research is `meta-prompting', where the language model is used to generate the prefix for the next generation \cite{reynolds_prompt_2021}. While we found that this produced intriguing results for our use case, for example by having the model first produce a list of types of people who interact with a topic, and then putting those phrases into a downstream template, we thought it added too much complexity.}

\subsection{Interface}

\begin{figure}
  \includegraphics[width=\textwidth]{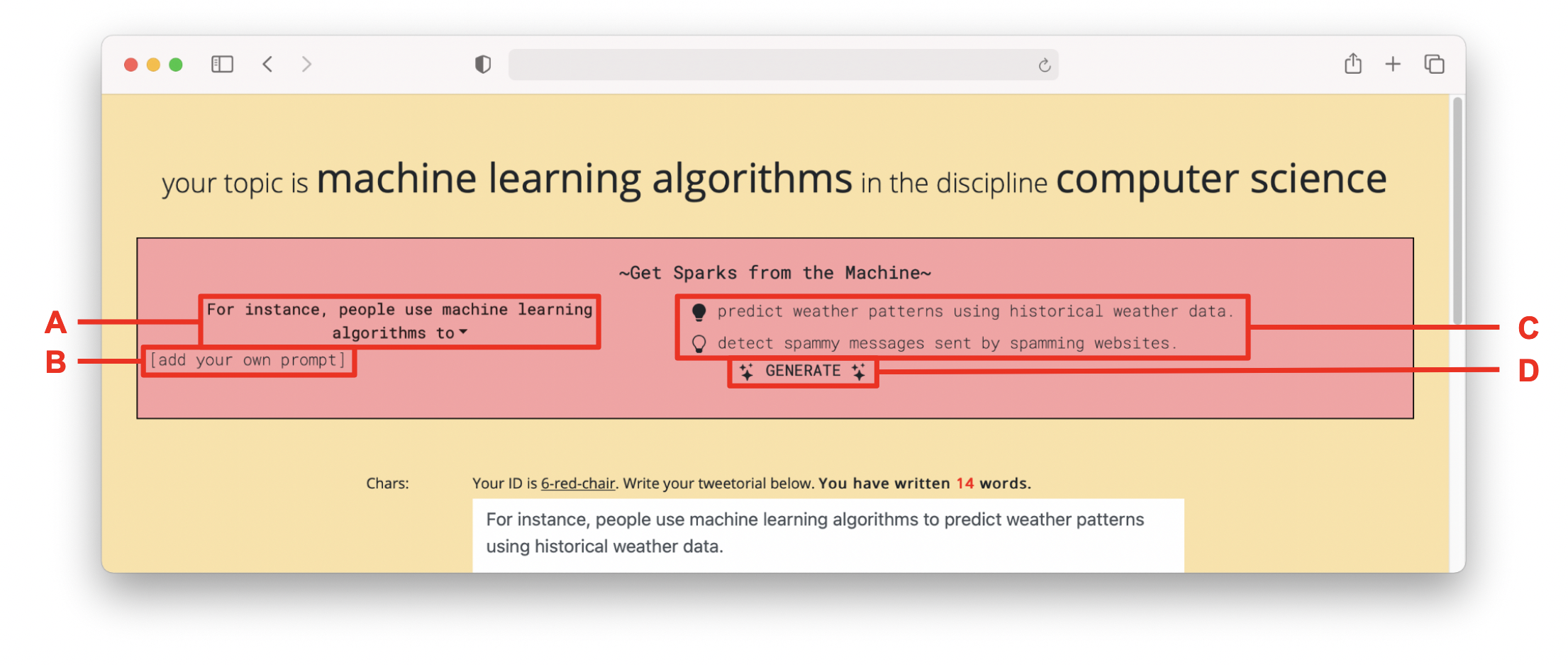}
  \caption{Example screenshot of our system generates sparks. A: writers can select from 10 template of prompts in a drop-down menu. B: writers can add their own prompt to the drop-down menu. C: sparks are generated with a lightbulb icon to the left, if writers click the lightbulb it will highlight and the spark is copied into the text area. D: writers can hit the generate button in order to generate a new spark.}
  \Description{adfsdfasdfsdfsad.}
  \label{fig:system}
\end{figure}

We design a website that takes in a writer's topic and context area. \autoref{fig:system} shows a screenshot of the system with its important features marked. The website consists of a single textbox for writing, and a `prompt box' above it that allows writers to interact with the sparks. Writers can select a templated prompt from a dropdown menu, or type in their own prompt and add it to the dropdown list. When a prompt is selected, if they press `GENERATE' the language model will generate a single spark. Writers can `star' a spark by clicking on the lightbulb icon -- this fills in the lightbulb and also pastes the spark into the textbox. If a writer selects a different prompt, the sparks already generated are preserved such that if they return to a previous prompt their generated sparks will be shown again.

\begin{figure}
  \includegraphics[width=.6\textwidth]{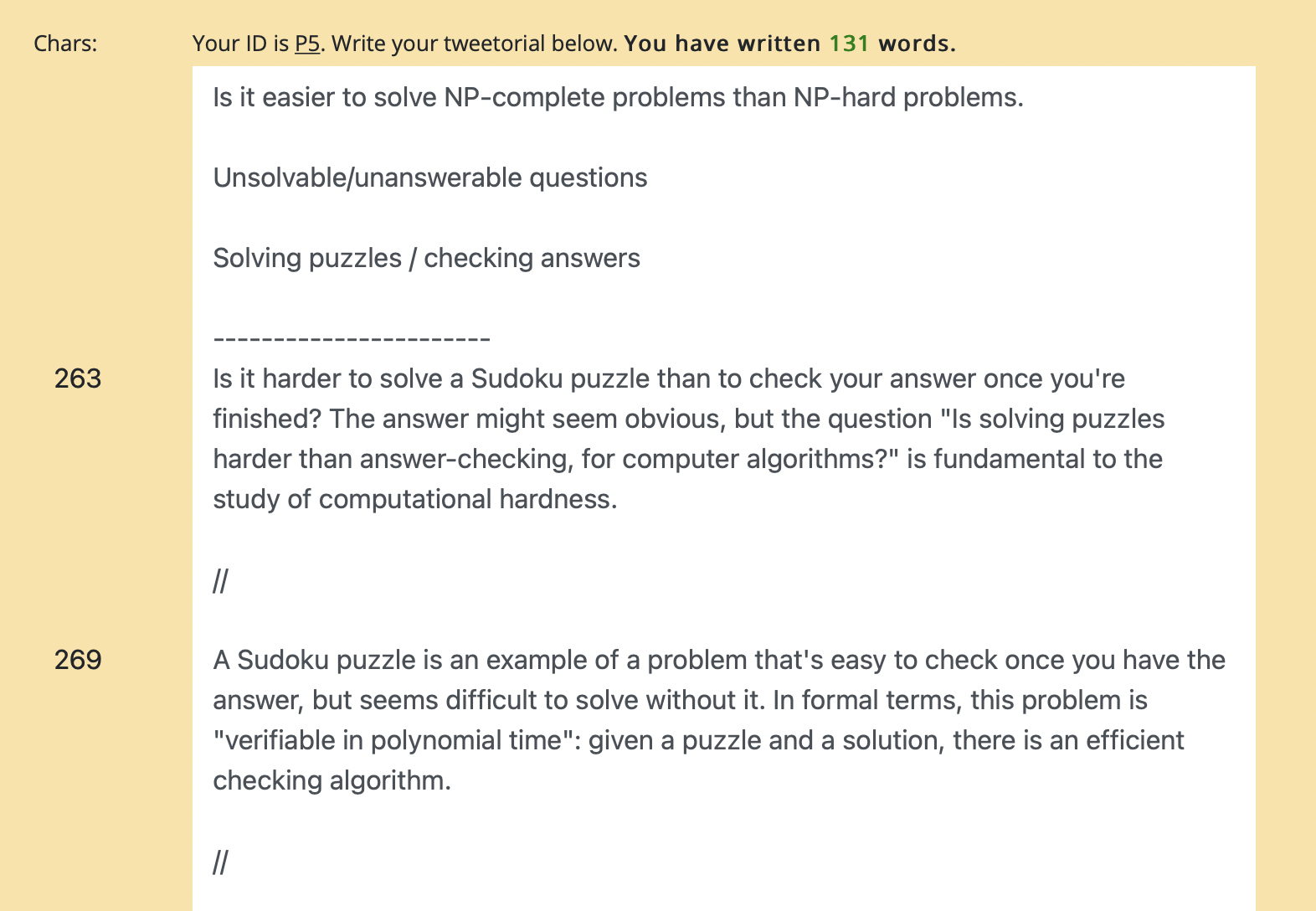}
  \caption{Screenshot of the text area from our user study. At the top is a word count, which counts only the words below the dashed line. Text above the dashed line is interpreted as brainstorming or notes. Participants can separate tweets with a double `//', and the character count for each tweet is shown to its left.}
  \Description{adfsdfasdfsdfsad.}
  \label{fig:textbox}
\end{figure}

The textbox contains some features useful for the tweetorial writing task. The textbox is split into two sections with a line of dashes. Above the line is reserved for brainstorming and notes, a feature writers requested and found useful during pilot studies. Below the line is the text area for the tweetorial writing. A word count for the writer's tweetorial draft is displayed at the top of the textbox, and a character count for each tweet (separated by line breaks and two forward slashes) is displayed to the left. \autoref{fig:textbox} shows these features with an example from our user study.

The website is implemented using Python 3.7 and the Flask web framework.\footnote{Link to demo to be released after anonymous review.}

\section{Study 1: Spark Quality}

We wanted to evaluate the quality of the sparks generated by our system. In particular, we wanted to evaluate how well the sparks, in isolation (i.e. not in a writing task), met our design goals of coherent and diverse. We also wanted to test how well the sparks could support a wide range of topics, and if certain prompts supported some topics better than others. To do so, we compared the sparks generated by the custom decoding method to a baseline system, as well as a human-created gold standard.

We have three hypotheses:

\begin{itemize}
    \item H1: The custom decoding produces more coherent and diverse outputs than a baseline system, but less coherent and diverse outputs than a human-created gold standard.
    \item H2: The custom decoding performs consistently across many different topics.
    \item H3: Some prompts work better for some topics.
\end{itemize}

\subsection{Methodology}

We wanted to evaluate the quality of ideas for a variety of topics. We selected three disciplines that have a glossary of terms page on Wikipedia, and that have been demonstrated to be a rich discipline for science writing on social media.\footnote{e.g. \url{https://twitter.com/dannydiekroeger/status/1281100866871648256}, \url{https://twitter.com/GeneticJen/status/897153589193441281}, and \url{https://twitter.com/meehancrist/status/1197527975379505152}} These disciplines were computer science, environmental science, and biology. For each discipline we randomly sampled 10 topics from their glossary of terms page. See the appendix for the full list of topics studied.

\subsubsection{Collecting a human-created gold standard.}

We wanted to collect human responses to our prompts to represent a gold standard or upper limit on the quality of ideas these prompts can generate. To do this, we recruited 2-3 PhD or senior undergraduate students in each discipline and had them complete the same prompts the language model did. Each student was paid \$20/hour for as long as it took them to finish the task.

We explained to them that the purpose of the prompts was to generate ideas to support an expert writing about the topic for a general audience. Each student had to complete 5 prompts per topic in 3 different ways, and was told to make the completions for a given prompt+topic combination maximally different. They were also instructed to ensure their completions were accurate, given their understanding of the topic, and that they could reference the web if they needed to check anything, as well as use web search results for inspiration. Finally, we explained that their ideas should be as concrete and specific as possible. Each student completed 5 prompts for the 10 topics in their discipline, for a total of 5 x 10 x 3 = 150 completions per person. It took them on average 3.5 hours  to come up with completions for all 10 topics in their discipline, and in the end we had 6 high quality completions per prompt+topic combination.

\subsubsection{Baseline language model condition.}

We compare the custom decoding to a language model baseline: group beam search with hamming diversity penalty. This is a strong baseline that encourages diversity in the way \cite{vijayakumar_diverse_2018} recommends, and can be implemented using arguments in the ‘generate’ function in the huggingface transformer library. Both the custom decoding and baseline model use the same underlying language model.

\subsubsection{Measuring coherence and diversity.}

Coherence is notoriously difficult to measure automatically, especially without training data -- measures like perplexity merely measure an output's likelihood under the model itself. For this reason we recruited 10 domain experts to annotate outputs for coherence on a 0 - 4 scale, in line with knowledge graph evaluations \cite{li_commonsense_2016}. For biology we had 3 senior undergraduate students majoring in biology; for environmental science we had 2 senior undergraduate students majoring in environmental science; for computer science we had 2 PhD students from the computer science department.\footnote{The students could not have also participated in the generation portion.} 
Each discipline had 900 sentences to annotate (300 human generated, 300 from the baseline model, and 300 from the custom decoding).
250 randomly selected outputs from each discipline were annotated by two different domain experts, and the Cohen's weighted kappa was calculated as: $\kappa=.54$ for biology, $\kappa=.51$ for environmental science, and $\kappa=3.4$ for computer science. Given that the agreement was moderate, we had a single annotation for the remaining sentences.

We measure diversity with sentence embeddings \cite{reimers_sentence-bert_2019}, in particular we report the average distance between outputs within a given prompt. A higher average distance means that outputs are more dissimilar, and therefore more diverse.

\subsection{Results}

\begin{figure}
     \centering
     \hfill
     \begin{subfigure}[b]{0.48\textwidth}
         \centering
         \includegraphics[width=\textwidth]{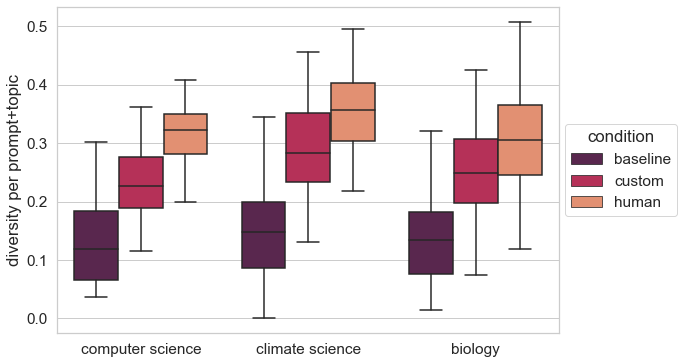}
         \caption{Distribution of diversity, split by discipline. Diversity is measured as the average sentence embedding distance per prompt+topic combination.}
         \label{fig:diversity}
     \end{subfigure}
     \hfill
     \begin{subfigure}[b]{0.48\textwidth}
         \centering
         \includegraphics[width=\textwidth]{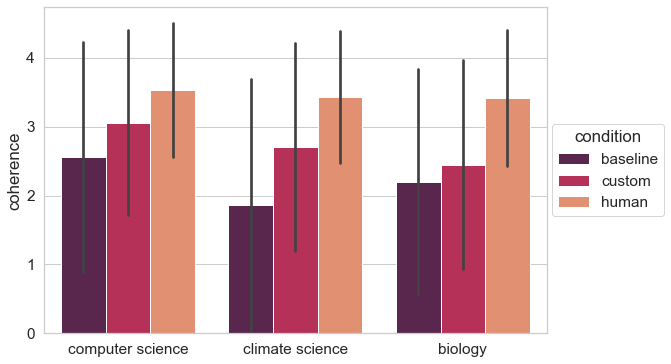}
         \caption{Mean coherence per prompt+topic combination, split by discipline. Each prompt completion was scored by a domain expert on a scale of 0 to 4.}
         \label{fig:coherence}
     \end{subfigure}
        \caption{Diversity and coherence measures across three test disciplines for three conditions: a baseline language model, a language model with the custom decoding, and a human-created gold standard. The custom decoding improves upon the baseline and approaches the human gold standard.}
        \label{fig:three graphs}
\end{figure}

Overall, the baseline had low diversity and coherence across all disciplines, while the human-created outputs perform much better. \autoref{fig:diversity} and \autoref{fig:coherence} show that the custom decoding method outperforms the baseline, but does not reach the performance of the human-created outputs. For diversity, two-tailed t-tests show this to be a significant difference for all disciplines (computer science: $p<.001$, climate science $p<.001$, biology: $p < .001$); for coherence, mann-whitney U tests show this to be a significant difference for all disciplines (computer science: $p<.001$, climate science $p<.001$, biology: $p < .001$).

\begin{table}
  
  \begin{tabular}{llll}
    \toprule
    condition & coh & div & One attribute of source code is...\\
    \midrule
     &&&it is typically written in a human-readable format.\\
    human created&4&.38& editability, so that programmers can easily change it to suit their needs.\\
    &&&it is a description a computer program.\\
    \hline
     &&&that it contains code written by humans.\\
    custom decoding & 4 &.37&its modularity - code modules contain reusable code components.\\
    &&& complexity.\\
    \hline
    &&&that it can be used as a source of information.\\
    baseline model & 2.6&.08&that it can be used as a source of inspiration.\\
    &&&its modularity.\\
    
  \bottomrule
\end{tabular}
\caption{Example outputs from our three conditions for a single prompt+topic combination, and the average coherence (coh) and diversity (div) scores for each set of three outputs.}
  \label{tab:outputs}
\end{table}

\autoref{tab:outputs} shows some example outputs from each conditions for a single prompt+topic. These examples demonstrate the quality of the human generated outputs: they are long, detailed, and diverse. Comparatively both language model methods are shorter, less specific, and more repetitive. However, the custom method seems to improve the overall quality of the outputs.

\begin{figure}
         \centering
         \includegraphics[width=.8\textwidth]{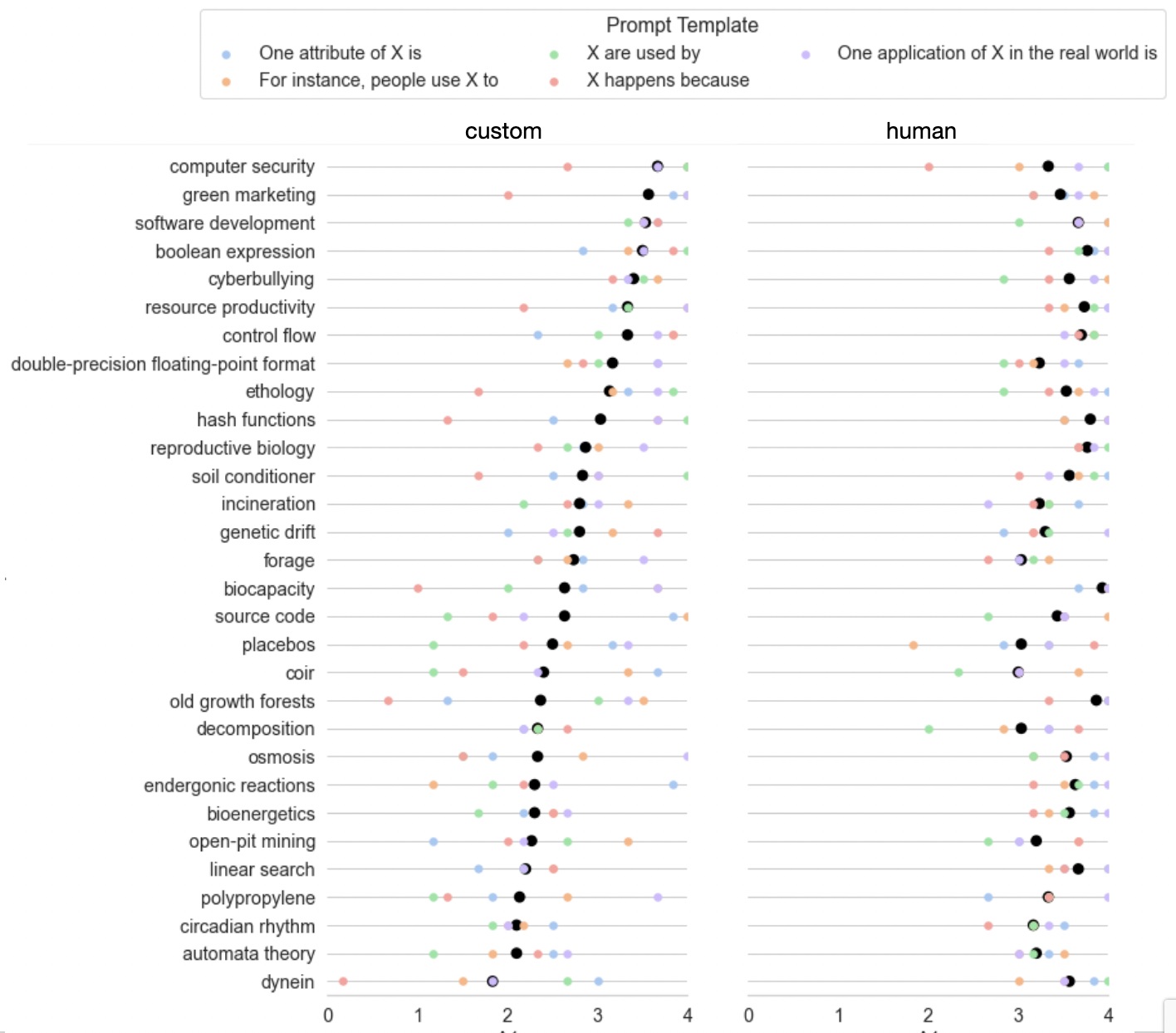}
        \caption{This graph shows the coherence per topic for the custom decoding and the human-created gold standard, where 0 is nonsensical or untrue and 4 is generally true. The black dot shows the average coherence of all responses for a given topic, while the colored dots show the average coherence for a given topic per prompt. Topics are ordered by average coherence in the custom decoding. This graph shows that some topics perform much better than others with custom decoding, while the human outputs are generally high quality regardless of topic. It also shows that within a topic there can be a large variation between prompt templates.}
        \label{fig:prompt-coherence}
\end{figure}

It is important to acknowledge that the variation in both the diversity and coherence measures are quite large. This means that while on average the custom decoding is an improvement over the baseline, and on average the human-created outputs are better than the language model outputs, for any given prompt+topic combination the output could be very high quality or of a much lower quality. People using the system will not necessarily see this huge variation; they will only see the 10 or so model outputs that they generate. 

For this reason, we dig into the variation by topic and prompt. \autoref{fig:prompt-coherence} shows the average coherence per topic for the custom decoding method and the human-created outputs. It plots the average coherence for each topic with the black dots, and the coherence for each prompt+topic combination in the colored dots. From this we can see that the variation in quality over the topics for the custom decoding method. For instance, the "computer security" outputs score an average of 3.7 in coherence, while "automata theory" outputs score 2.1. When looking at the human-created outputs, the quality is far more consistent, with no topics dropping below an average of 3 in coherence. 

This demonstrates that our system works well for some topics and less well for others. While we expected that our system would not perform as well as a human would, we did expect that the system would perform more consistently across topics. It is unclear why the language model performs significantly better on some topics, and given the way that these language models are trained it is difficult to inspect or even predict how well the model will perform on a given topic.

\autoref{fig:prompt-coherence} also shows that some prompt templates work better for some topics than others. In our system, the quality of outputs vary significantly with the prompt template. In the human generated outputs, the variation is smaller, but still we see some range. For instance, let's look at the topic "dynein", the worst performing topic. The prompt "Dynein happens because" scores an almost 0 on the 0 to 4 coherence scale, while the prompt "One attribute of dynein is" scores a 3. Dynein is a family of proteins important in cell behavior. Given this, it makes sense that the system is more likely to produce coherent outputs on attributes, rather than why this family of proteins "happens".

However, it's notable that the human outputs scored 3 or above for all prompts for "dynein". Here is a human output about why dynein happens: "Dynein happens because organelles, such as the Golgi complex, need to be positioned in cells." Though this sentence structure is a little convoluted, it's clear that the human was able to compensate for the prompt and still write something coherent and meaningful.

Seeing the difference a prompt template can make highlights the importance of using a prompt that works well for the topic. Since we wanted to test our system with unseen topics, we ensure that participants can add their own prompts in case the templated prompts don't work well for their topic.

\section{Study 2: User Evaluation}

We evaluated how our system supported PhD students in writing tweetorials. Tweetorials are short explanations of around 500 words which have a low-barrier to entry and are gaining popularity as a science writing medium \cite{soragni_scientists_2019}.  We use PhD students as they are eager to participate in science writing \cite{howell_engagement_2019} and many tweetorials are already written by PhD students, demonstrating that this a writing task our participants may conceivably want to engage in on their own. This study was approved by the relevant IRB.

\subsection{Methodology}

We recruited 13 participants, all students from five different STEM disciplines, to write tweetorials on a topic of their own choice, related to their area of study. By letting the participants pick their own topic, we ensured that they were writing within their area of expertise, and we were able to test our system on unseen topics.

Participants were first asked to read an introduction to tweetorials, which explained what tweetorials are and walked through an example tweetorial. They were then introduced to the system and watched a short video that demonstrated the system's features, and showed an example use case of the system when writing about 'machine learning algorithms'. Participants could ask clarifying questions to the facilitator. If participants asked to learn more about how the system worked, the facilitator said that it was an algorithm that could generate text in response to a prompt, and that they could discuss the system further after they completed the writing task.

At this point the participant was asked to pick a topic to write about, as well as provide a `context area' to aid the system in correctly interpreting their topic. Then they were given 15 - 20 minutes to interact with the system and write approximately the first 100 words of their tweetorial. Mouse clicks and key presses while the participant interacted with the system were collected, as well as all sparks generated.

After this, the participant filled out a short survey and partook in a semi-structured interview with the facilitator. The survey questions and the questions that structured the interview can be found in the appendix. The study took about an hour and participants were compensated \$40 USD for their time.

Participant interviews were transcribed and the authors performed a thematic analysis \cite{cooper_thematic_2012} on the interview transcripts. The analysis centered on three areas: how sparks were helpful, how sparks were unhelpful, and ownership concerns in response to writing with a machine. Relevant quotes were selected from the transcripts and collated in a shared document, where is the author discussed and collected the quotes into emergent themes.

\subsection{Results}

\begin{figure}
  \includegraphics[width=\textwidth]{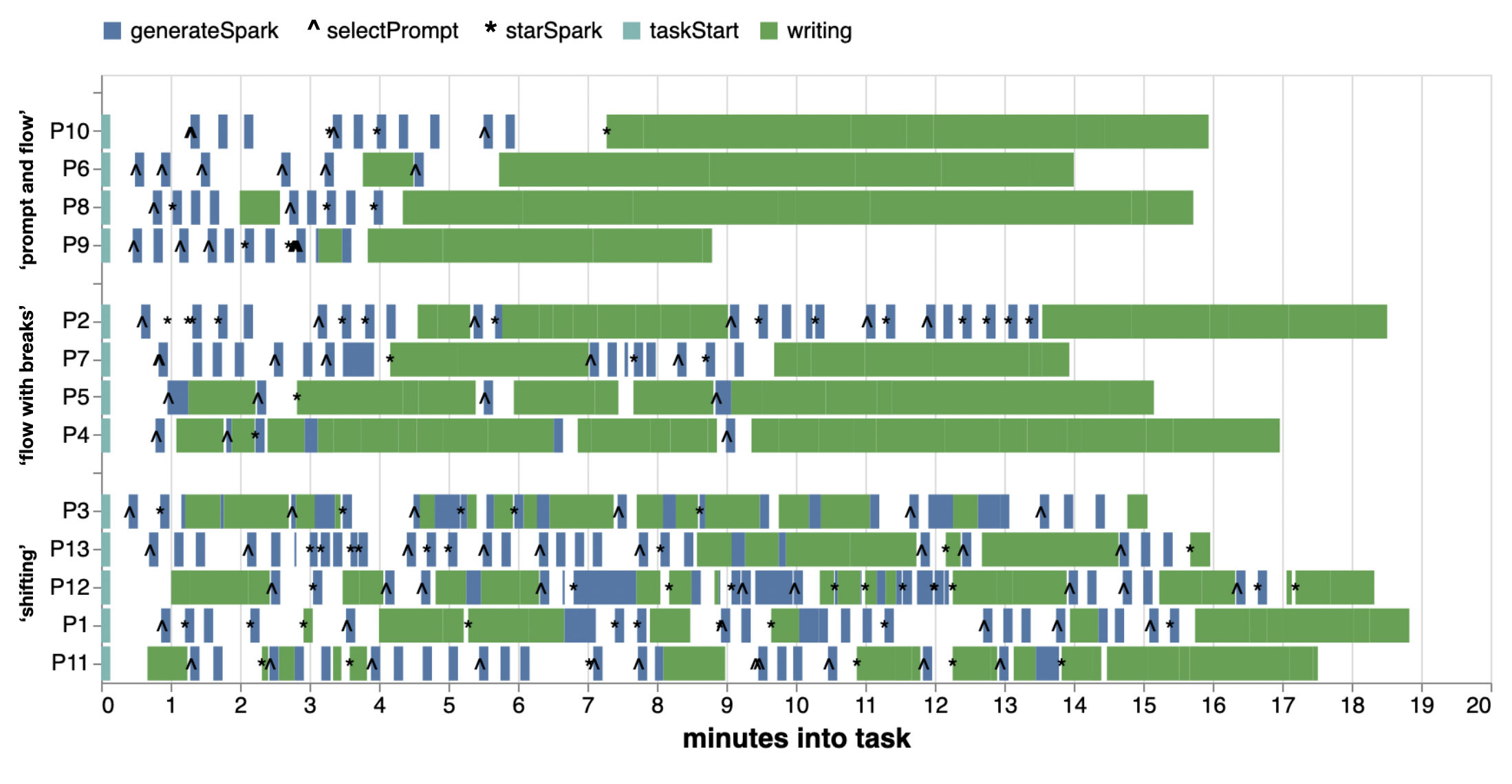}
  \caption{Timelines of all participants from the study, with time writing versus time generating sparks marked in different colors. Participants are grouped by their engagement pattern.}
  \Description{adfsdfasdfsdfsad.}
  \label{fig:timeline}
\end{figure}

We report on participant demographics and topic selection in \autoref{tab:participants}, as well as the prominent themes that emerged through our analysis in \autoref{tab:themes}. Our thematic analysis covered three main areas: ways in which sparks were helpful, ways in which sparks were unhelpful, and how participants felt about incorporating sparks into their writing. For each area, we report themes with prevalence greater than 20\%. Prevalence is measured by the number of participants who brought up that theme in their interview. 
In the case of ownership, there was a very high variability in responses, such that no one theme had over 20\%. For this reason we report the reasons that people brought up in the text, but do not list them in the table. Behavioral data, such as time spent writing and how many sparks were starred, are also reported and timelines for each participants' activity can be found in \autoref{fig:timeline}.

\begin{table}
  \caption{Participant Demographics. Low = once a year or so. Med = Once a month or so. High = once a week or so.}
  \label{tab:participants}
  \begin{tabular}{ccC{2.3cm}cc}
    \toprule
    ID & Discipline    & Science Writing (general / twitter)&Topic&Context Area\\
    \midrule
P1   &    Climate Science &    Low / Low &              rainfall variability  &              climate science \\
P2   &    Climate Science &    Low / Never &          predicting climate change &               climate science\\
P3   &    Climate Science &   Never / High &                   sea level change &                    geophysics\\
P4   &    Climate Science &    Low / Low &  glacier retreat over the holocene &                  paleoclimate\\
P5  &   Computer Science &   Low / Never &      computationally hard problems &              computer science\\
P6  &   Computer Science &    Never / Never &                   pseudorandomness &  theoretical computer science\\
P7   &  Political Science &   Med / Med &                document embeddings &   natural language processing\\
P8   &         Psychology &               Never / Low &                     regulatory fit &                    psychology\\
P9  &         Psychology &  Low / Low &      motivated impression updating &             social psychology\\
P10   &      Public Health &    Low / Low &              measurement of sexism &                     sociology\\
P11   &      Public Health &               Never / Never &                logistic regression &                  epidemiology\\
P12   &      Public Health &      Low / Never &                deprivation indices &                 public health\\
P13   &      Public Health &     Med / Med &                  threat multiplier &          environmental health\\
  \bottomrule
\end{tabular}
\end{table}

\renewcommand{\arraystretch}{1.25}
\begin{table}
  \caption{Thematic analysis}
  \label{tab:themes}
  \begin{tabular}{|p{4.5cm}cp{8cm}|}
    \hline
    \textbf{Code} & \textbf{Prevalence} & \textbf{Example Quote}\\
    \hline
    \multicolumn{3}{c}{\textit{reasons sparks were helpful}}\\
    \hline
    Crafted concise, detailed sentences. &54\%&
    Most of the time it [the system] was articulating the ideas that were already in my head in a way that's short and concise.\\
    Came up with ideas or angles. &46\%&
    It [the system] reminded me, Oh, it's not just my application, there's these other people using the same technology, but working on other problems. \\
    Showed reader perspectives. &31\%&
    It [the system] reminded me that there might be a more common understanding of this thing that I'm writing about, that's different from the highly specific one I've been living in.\\
    \hline
    \multicolumn{3}{c}{\textit{reasons sparks were unhelpful}}\\
    \hline
    Incorrectly interpreted the topic. &38\%&
    It [the spark] just wasn't helpful, but only because it was using the different sense of `embedded'.\\
    Inaccuracies.&23\%&
    Some of the sparks said, like, logistic regressions are used to estimate relative risks, which is completely not true.\\
    Not desired angle.&23\%&
    Someone probably does really care about measuring sexist attitudes ... but it just isn't my focus.\\
    Vagueness.&23\%&
    I would say about 20\% of them were just not specific enough to warrant talking about.\\
    \hline
\end{tabular}
\end{table}

\subsubsection{Participant demographics and topic selection.}

The 13 participants came from five STEM disciplines, with the most common disciplines being Climate Science and Public Health. All but one were doing PhD (the remaining doing a research Master's) and varied from their 2nd year to their 7th year in their program. Participants were asked how often they wrote about technical topics for a public audience, and how often they did so on Twitter. Most participants rarely or never did so, though a few did so on a monthly or even weekly basis. 

Participants were asked to select a topic they understood well that was related to their research. The facilitator attempted to aid participants in selecting a topic that wasn't too broad, but also not too specific, but as the facilitator did not necessarily have the same expertise as the participant this was at times difficult. Participants selected a wide range of topics, with no overlap. The full demographics and topic selection can be found in \autoref{tab:participants}.

Given the diversity in the participants' topics, how well the system generated sparks on their topics, and how they articulated or responded to questions in the interview, there was a high variability in how participants felt about the system. For this reason, a prevalence of 50\% or above is considered very high. This would mean that over 50\% of participants independently responded in the same way to an open ended question, despite writing about a unique topic and seeing a unique response from the system.

\subsubsection{Participants had different engagement patterns when interacting with the system.}

Overall we saw many different ways in which participants interacted with the system. Some participants generated and starred many sparks, while others generated only a handful and starred two or three.
The average number of sparks generated was 17.2 (std=10) and the average number `starred' was 5.1 (std=4.3). Additionally, it did not seem that generating or starring more sparks necessarily meant a participant found the system more helpful. Some participants who starred just two or three sparks talked at length in their interview about how useful the system was. Others who starred more than 10 complained that the system produced only vague or inaccurate outputs.

However, we did see themes of how people tended to interact with the system.
\autoref{fig:timeline} shows all participants' task timeline, split into three archetypal patterns: `prompt and flow', `flow with breaks', and 'shifting'. We talk about each in detail:

\textbf{`Prompt and flow'} participants look at the sparks at the beginning, before they start writing, and then spend the rest of their time writing, never going back to generate more sparks. This kind of participant used ideas to jump start the writing, but once they were writing had no need to return to the sparks. Several participants described getting some kind of inspiration from the sparks and then "flowing through an argument". Once these participants got into the `flow' of writing, they didn't need any further support.

Although these participants don't generate more sparks, they may be using several sparks that they had `starred' earlier. For instance, here are the first three tweets written by P8, where  sentences or phrases inspired or highlighted:

\begin{quote}
\small
    Have you ever wondered why certain activities just feel right to you? Like you are able to do well and feel good in the process? \hl{Research on regulatory fit helps explain why certain activities feel right versus wrong...} (1/n)

    //

    \hl{Regulatory fit happens because people can be driven by a range of different fundamental motives, and distinct activities "fit" each of these motives.} For instance, people who are most motivated by their duties and obligations to others prefer activities that (2/n)

    //

    Involve carefully defending and protecting what they find important and methodically scrutinizing information. In contrast, people who are most motivated by hopes and aspirations prefer activities that  involve eagerly making progress and engaging in curious exploration. (3/n)
\end{quote}

Although they wrote all of this without generating more sparks, they did make use of two sparks -- one for the first tweet (``Research on regulatory fit reveals that regulatory fit varies considerably across individuals depending upon factors.") and another for the second (``Regulatory fit happens because individuals behave differently under varying conditions."). So while they did not return to generating more sparks, they did return to the sparks they had previously starred.

\textbf{`Flow with breaks'} style participants are similar, but they pause in the middle of the writing and go back to sparks. This type of participant talked about returning to the sparks when they "would get stuck" or at some kind of inflection point in the writing process. P2 described this as, "there are a bunch of ideas that I need to kind of weave together before using the tool again, when I get more stuck." In both of these types of interactions, we see that the writer is using the sparks to jump-start their writing whenever something makes them pause, but when they are writing they are writing continuously, even if they are drawing on sparks they had earlier selected as potentially useful.

\textbf{`Shifting'} style participants are often moving between writing and generating sparks. They are constantly returning to the sparks for different details, or for help with crafting a sentence. They tended to star a lot of sparks and use a lot of custom prompts, as what they were looking for from the sparks changed as they proceeded through their writing, and a lot of language in their writing draws directly from a spark. For example, consider the start of P1's tweetorial on rainfall variability, where sentences or phrases inspired by sparks are highlighted:

\begin{quote}
\small
    Do you like coffee?? 

//

Want to understand how coffee and climate change are related? 

//

\hl{Rainfall variability happens because rain falls differently depending on temperature and humidity.}

//

\hl{Since temperature and humidity vary around the globe, we observe different rainfall patterns across the world.} 

//

\hl{For example, rainfall variability may cause extreme precipitation events leading to flooding or a lack of precipitation will lead to drought.}
\end{quote}

The last three tweets shown here are all taken directly from sparks, with minimal changes. It almost seems like collage, where the writer is mostly arranging language that came from the system.

This suggests there are different engagement patterns that people use when interacting with machine-generated suggestions. Although we expected participants would mostly use sparks for the beginning of their tweetorial, we found that participants used sparks at all points of their writing process. In the next sections, we report on the various ways in which participants made use of the sparks.

\subsubsection{Sparks helped participants craft concise and detailed sentences quickly.}

Although we intended the system to inspire participants with new ideas, the most prevalent reason the participants cited for the spark being helpful was for crafting sentences. Many participants remarked that although the sparks were showing them information that they already knew well, it was much faster and easier to draw on language from the sparks than to write a sentence from scratch.

For instance, P12 said:

\begin{quote}
    Most of the time [the system] was articulating the ideas that were already in my head in a way that's short and concise, which is useful. Like `deprivation index measures the relative deprivation experienced by an individual relative to others,' that would have probably taken me like three sentences to write, then I'd have to spend time editing it down. And then yeah... this is a lot quicker.
\end{quote}

\begin{figure}
     \centering
     \begin{subfigure}[b]{0.49\textwidth}
         \centering
         \includegraphics[width=\textwidth]{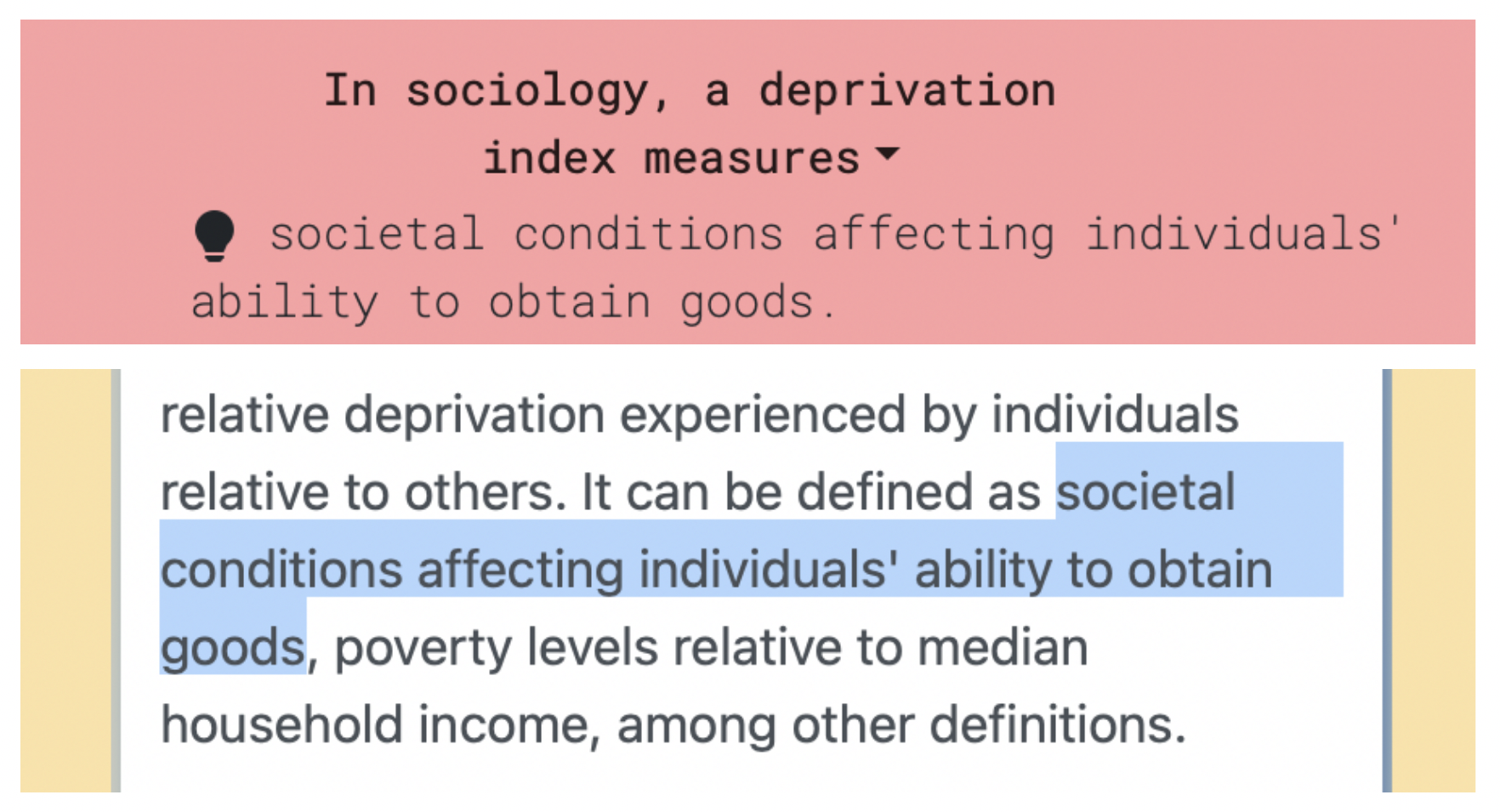}
         \caption{An example of how a participant in our study used a spark for crafting a detailed sentence. Highlighted text was inspired by the spark.}
         \label{fig:ex1}
     \end{subfigure}
     \hfill
     \begin{subfigure}[b]{0.49\textwidth}
         \centering
         \includegraphics[width=\textwidth]{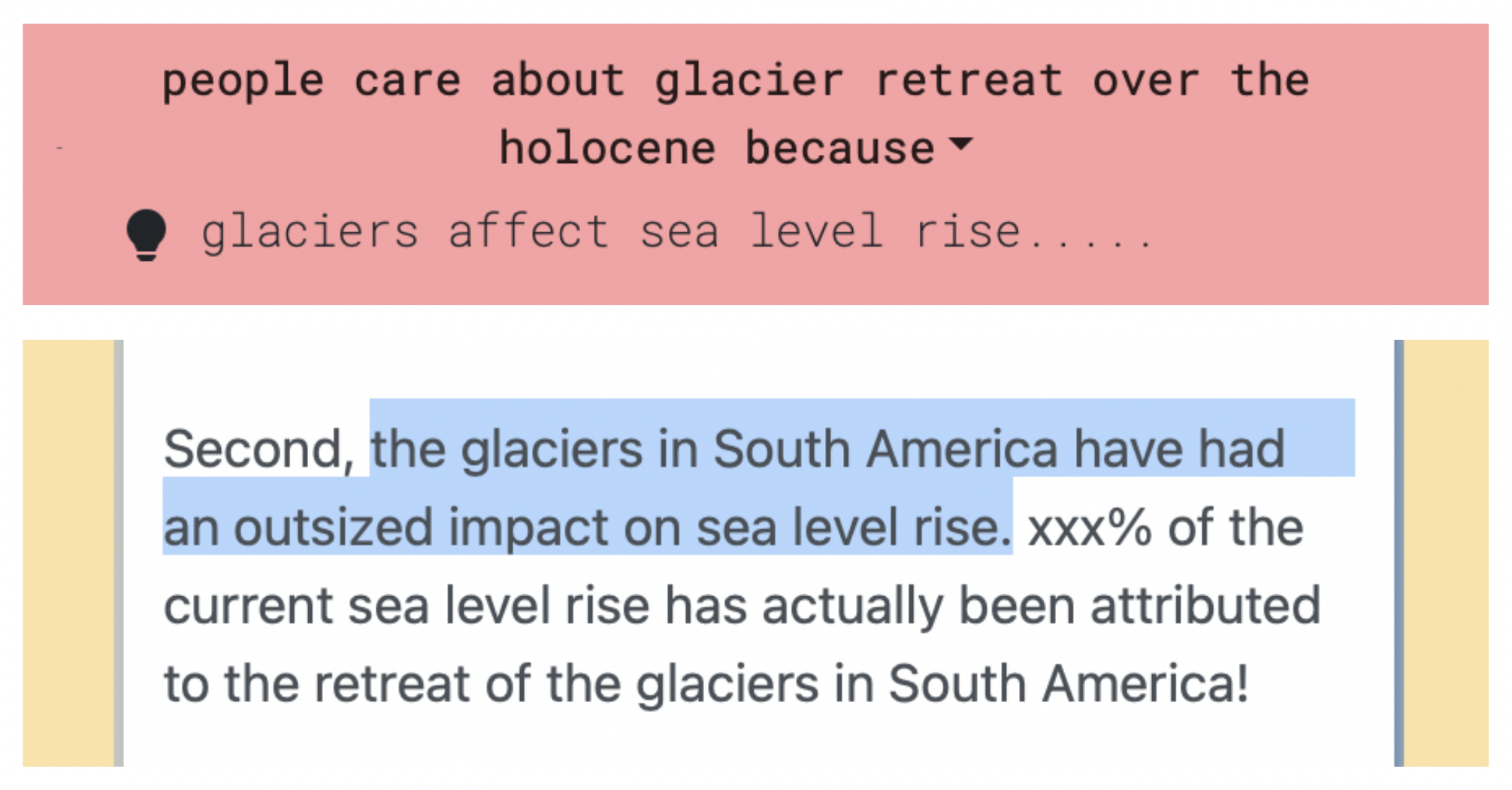}
         \caption{An example of how a participant in our study used a spark to make their tweetorial more engaging to a general audience. Highlighted text was inspired by the spark.}
         \label{fig:ex2}
     \end{subfigure}
     \hfill
\end{figure}

\autoref{fig:ex1} shows another example in which P12 drew on a spark in order to write a clear and concise definition.

Several participants noted that they often go to Google or Wikipedia simply to get a well-written definition of a topic they understand well. This is something that the system was able to do for them without requiring a click away from the writing interface and incurring a change in context. P7 noted that it did a good job compressing what he would have looked for or found a Google search. P8 noted that all the sparks were similar to what she would have found on Wikipedia or via a Google search, but that they were ``bite-sized" or ``sound bite ready".

\subsubsection{Sparks reminded participants of other ideas or angles about their topic.}

Several participants noted that the sparks provided good ideas or angles for discussing or introducing their topic. P2 noted that `weather prediction models' were useful entry point to their research. They said, ``that's something within my field that the general public might be more familiar with than what I actually do.'' 

\autoref{fig:ex2} shows how P4 drew on a spark about the `sea level rise' in order to make their topic of `glacier retreat over the holocene' interesting to the reader. P4 said in their interview, ``It's often hard to figure out how to spin things in ways that feel relevant to people who don't study this," and that the sparks helped her find ways to make her research relevant. P7 said, ``[the system] definitely generated multiple [ideas] that I could have written different tweetorials about."

When asked if the sparks were giving them new ideas, many participants said that the system was helping them get to ideas they likely would have come up with themselves, but faster. For instance P4 said, ``It was definitely faster. I think I would have gotten there, but it would have taken me longer."

\subsubsection{Sparks encouraged participants to think about common reader perspectives.}

Several participants noted that the sparks reminded them of how people reading their tweetorial might be interpreting their topic. For instance P10, who was writing about measuring sexism, noted that many of the sparks talked about sexist attitudes. She said the while that certainly is an aspect of measuring sexism, it isn't the aspect that she actually studies. And so the sparks reminded her that people's main assumptions when thinking about sexism is probably about attitudes and therefore that might be an assumption that she will have to address in her tweetorial.

P5, writing about computationally hard problems, noted that one of the sparks talked about NP-completeness. He said that while at first he thought this might be too technical, it then made him wonder if someone who was reading a tweetorial about computational hard problems might already know at least some of the keywords about this topic. In this way the sparks made him reflect on what knowledge his readers might already have.

\subsubsection{Sparks failed in different ways for different participants.}

The ways in which participants found sparks to be unhelpful varied highly. The most common reason participants said sparks were unhelpful was that they incorrectly interpreted their topic. In this case, the sparks were not necessarily incorrect, but rather they reflected some alternate interpretation of their topic. For instance, P12, who was writing about deprivation indices, said that some of the sparks were about obesity. Obesity has little to do with deprivation indices, but they thought the algorithm may have been associating deprivation with nutrition. Similarly P8, who was writing about regulatory fit, commented on several sparks about government regulation, which is unrelated to her psychology topic, but she assumed the algorithm was simply free associating with the word `regulatory'.

Other reasons participants found the sparks to be unhelpful were factual inaccuracies, dealing with aspects of their topic that they were not trying to explain or that they did not study, and vague outputs. Participants also mentioned that some sparks were nonsensical, tautological, had too much jargon, or were simply "bizarre".

Overall participants varied highly in how useful they found the sparks. Some participants found that the sparks were so low quality that they found the system completely unhelpful. Others said that even though some of the sparks were not helpful the ones that were helpful were so helpful they were unconcerned with a few that didn't make sense or were off-topic. 

\subsubsection{While most participants had no ownership concerns for science writing, a few felt apprehensive about accepting machine-generated ideas.}

Most participants had no concerns about incorporating sparks into their writing. Several participants cited being very familiar with the material as a reason for not being concerned. Others said that since they are writing about public knowledge, it was unimportant where their ideas came from. One participant articulated that coming up with ideas is not the hardest part of science writing, but rather putting in the time and energy into building an audience and writing something engaging, so incorporating sparks would simply be one small part of a much larger endeavor that she took on. One participant compared the sparks to searching on Google; another compared it to Grammarly (a grammar-checking service). One participant said that the sparks were simply elaborating on his own idea, and thus he still felt ownership of the material.

A few participants did have ownership concerns. One participant talked about how he considered outreach and science writing to be part of his job as an academic, and thus any system that automated some aspect of this felt like it was taking over something that he found fundamental to his work. Several participants noted that machine writing was simply the future, and while they may have some apprehensions, they knew they would have to get over it.

Across all participants, whether or not they had ownership concerns, they wanted to ensure that anything they wrote was in their own voice and that they did not plagiarize. Several participants brought up that they were unsure exactly where the sparks were coming from, and they wanted to make sure that anything they took from the sparks was adequately changed, to alleviate any concerns about plagiarism. 
P2 described this as, "I think if I was using something like this, I would probably never use an entire sentence verbatim. Just because, if you don't know where it's pulling it from... I wouldn't want to run the risk of plagiarizing something accidentally even."
When asked if these edits would make the sparks less helpful, for instance if it meant it would take more time in order to edit what they chose to incorporate, most participants said that sparks would still overall speed up the writing process. But not everyone agreed; some participants thought the need to edit the sparks made them less helpful.

\subsubsection{Custom prompts allowed participants to iterate on how they interacted with the system.}

Many participants noted that they were unsure how to get the best sparks out of the system. Several compared the system to Google but said it was easier to navigate Google because they already knew what kind of queries would get the best results. In the study though, they noted that they were unsure which prompts would produce the best sparks. Several participants even said that some of the sparks that were unhelpful may have been because their custom prompts didn't properly prime the system.

One way participants used the custom prompts was to get the system to better understand their topic.
P13's discussed how the model didn't seem to correctly interpret his topic of "threat multiplier". You can see him iterating on custom prompts in an attempt to produce more accurate results: 

\begin{itemize}
    \item ``threat multipliers are''
	\item ``climate change is considered a threat multiplier because''
	 \item ``climate change is considered a threat multiplier for health disparities because''
\end{itemize}

Participants also used the prompts to generate more specific sparks.
In the case of P7, the system seemed to understand his topic "document embeddings" well enough, but he was curious about specific aspects of this topic. You can see him looking for these in the following custom prompts:

\begin{itemize}
    \item ``Document embeddings help researchers''
	\item ``Document embeddings help search engines''
	\item ``Linguists can use document embeddings to''
\end{itemize}

Finally, others participants entered custom prompts that seemed to be used less for idea generation, and more as an auto complete. For instance, P12 entered several custom props designed to generate a definition:

\begin{itemize}
    \item ``The Townsend Index is''
	\item ``In sociology, a deprivation index measures''
	\item ``the variables in the Townsend Index are''
\end{itemize}

Overall, we saw many different uses of the custom prompts, and found them to be an important aspect of the system. 

\section{Discussion}

\subsection{How to best use large language models in constrained writing tasks.}

Creativity requires both convergent and divergent thinking \cite{taura_creativity_2011}. We expected that the sparks would support participants with divergent thinking, by showing them how their topic might connect to things of interest to readers. And indeed many participants cited this as a main use case of the sparks. However, we found that many participants also found that sparks helped with convergent thinking. Sparks helped participants put the implicit, detailed, and often messy knowledge they had in their mind into concise sentences.

Existing work on large language models for storytelling tend to focus on divergent thinking. Language models provide writers with new ideas for plot points, character attributes, scene descriptions, and the like \cite{sudowrite, aidungeon}. However, what has typically been ignored is the convergent aspect of storytelling, for instance the tying together of plot points into a satisfactory ending. This is likely because of the difficulty language models have had in the past with staying on topic, though this may be changing as language models improve.

One of the reasons that the sparks may have been able to help participants with convergent thinking is because participants already knew what they were trying to explain. If we think of storytelling in response to a writing prompt as starting somewhere and perhaps not knowing the destination, we can think of writing tweetorials as knowing the destination but not where to start. In this case, participants were able to look at the sparks and recognize some of them as a convergence of the ideas already in their mind. 

We sometimes saw this when participants would enter custom prompts into the system. They often knew the correct way to complete the prompt, but didn't necessarily know the best phrasing. For instance, P12 entered in the prompts "The Townsend Index is" and "the variables in the Townsend Index are" not because they didn't know what the Townsend Index is or what its variables are, but because they wanted the system to provide a well-crafted sentence for them.\footnote{The authors of this paper did not know what the Townsend Index is. For the curious, it is a measure of material deprivation levels in a population, and it's variables are: unemployment, non–car ownership, non-home ownership, and housing overcrowding.} In this instance we might think of the language model as performing convergent thinking for the writer and letting the writer recognize what is correct and well-phrased.

An area of future development we see is more explicitly using the language model to help writers with different aspects of the writing process, for instance by developing prompts intended to directly spark new ideas and other prompts to return concise definitions or summaries. We are seeing this in the storytelling space, where systems are being developed to prompt plot ideas or specific scene descriptions or focus on rewriting in a different style.

\subsection{Three areas for technical development: better controls, increased breath, and graceful failing.}

However, to support science writing we will need to improve the underlying technology. We see three main areas of technical development that may be fruitful.

The first is improving the controllability of language models. While it is clear that pre-trained language models can generate sentences that span the spectrums of specificity and complexity and take on a number of styles and opinions, it is less clear how to ensure that a language model generates the kind of outputs a writer wants. Prompt engineering seems like a promising avenue for this, especially as prompt engineering can be done in natural language. If prompt engineering is done with vectors, it will be completely opaque to users, but if we are able to control language models through natural language prompts alone, users will have some intuitive understanding as to how to control the language model and are more likely to be able to use this control themselves.

The second is increasing the breadth of a language model's knowledge. Again, it is clear that pre-trained language models contain much world knowledge, but it is also clear that this knowledge is not evenly distributed. There have been several calls to be more careful and explicit with the data that these pre-trained models are trained on \cite{bender_dangers_2021, bandy_addressing_2021, caswell_quality_2021}. If we better understood the training data, then we might better understand how much a model can be reasonably expected to know and can better augment the training data to increase a model's breadth.

Third, language models do not always fail gracefully or in a way that a writer might understand. In our study, when reflecting on sparks from a custom prompt that were tautological or redundant, P1 said, "maybe it's that either way you slice it `atmospheric circulation patterns' is not the best way to start when you're talking to a general audience." In this case the participant was trying to make sense of why their custom prompt was producing low quality outputs -- sentences that were syntactically correct but contained no real knowledge. They assumed that if the model was unable to generate a reasonable sentence then perhaps the topic was too specialized for her audience. But we are not sure why the system produced redundant results; perhaps the prompt wasn't engineered properly or something else went wrong. Studies could be done on a model's knowledge of the training data to see how a model responds differently when prompted with topics it has varying levels of knowledge of. Then we might be able to methodologically design methods that detect why model produce certain low quality outputs. 

\subsection{More experience with language models may improve results for users.}

When asked how the system compared to using a web search engine, many participants noted that they had so much experience with web search that they were more comfortable navigating it to find what they needed. In contrast, when using the system, they were unsure what kinds of prompts would produce the results they wanted. How users craft web search queries, and how they overall use web search engines for information retrieval, is a long and active area of study \cite{yamamoto_exploring_2018, smith_domain-independent_2017}. For instance, advanced users query less frequently in a session, compose longer queries, and click on results further down on the list \cite{white_investigating_2007}.

It's possible that some of the issues participants encountered, for instance outputs that incorrectly interpreted the topic or did not reflect participants desired angle, could be improved if participants had more experience and a better understanding of how to prompt language models. For example, communities interested in text-to-image generative models have been crowd-sourcing "tricks" (such as the appending of a phrase) that help steer generations to their liking. People are discovering workarounds by probing the system, in a way that is separate from the research on better prompt engineering. They are learning how to use the tool and its depth of understanding to their full advantage. No amount of prompt engineering will remove the need for user input, in whatever form that may be, just like no matter how much work is put into delivering better search results, users still have to learn how to craft good queries and build expertise and understanding of the system.



\subsection{Limitations: the effects of personality, writing style, topic, and skill.}

Our user study was intended to be exploratory. The 13 participants came to the writing task with very different levels of experience in science writing, and very different topics for the system to respond to. Future work, and more developed systems, may be able to study these differences more carefully, perhaps with a larger set of participants, such that the impact of things like writing style or topic granularity may be able to be taken into consideration.

Additionally, it was clear that some participants were far more forgiving of system than others. For instance, some participants saw one or two inaccurate or redundant sparks and then considered that prompt or even the whole system to be a "dead end". Others were not phased by seeing sparks that didn't meet their needs. They even found ways to re-purpose unhelpful sparks to be useful, for example by interpreting inaccurate sparks as an indication that their topic may have been too niche or as a signal for how readers might also misinterpret their topic. Work on how users develop mental models of AI systems has shown that some people are more likely to blame the system when something goes wrong, and others are more likely to blame themselves \cite{gero_mental_2020}. Future work may want to study if personality type is a predictor of how useful participants find these kinds of systems, such that it does not remain a confounding factor.

Finally, our system uses just one particular language model with one particular decoding method. Improving the generative abilities of language models is a large and growing area work. We expect that their abilities will continue to improve, and this will obviously change the ways in which people react to them.

\section{Conclusion}

In this work we investigated how to use a large language model to support writers in the creative but constrained task of science writing. We developed a system that generates ``sparks'', sentences about a scientific concept intended to inspire writers. We found that our sparks were higher quality than a baseline system, and approached a human-created gold standard. We also found in an exploratory study with 13 PhD students that participants used the sparks in many different ways: to craft detailed sentences, to get ideas for how to engage the reader, and to better understand common reader perspectives. Finally, we discussed how language models might be used as support tools for writing in the future, what areas of technical development we believe will be fruitful, and how users might learn to interact with language models in a writing context.


\bibliographystyle{ACM-Reference-Format}
\bibliography{citations}

\appendix

\section{Methods for Study 1}

\subsection{Full List of Topics Studied}

\begin{itemize}
    \item \textbf{Biology}:
endergonic reactions, 
genetic drift,
decomposition,
dynein,
circadian rhythm,
placebos,
ethology,
osmosis,
reproductive biology,
bioenergetics.\\
Topics randomly sampled from  \url{https://en.wikipedia.org/wiki/Glossary_of_biology}.
\item \textbf{Environmental science}:
biocapacity,
resource productivity,
forage,
polypropylene,
open-pit mining,
soil conditioner,
incineration,
green marketing,
coir,
old growth forests.\\
Topics randomly sampled from
\url{https://en.wikipedia.org/wiki/Glossary_of_computer_science}.
\item \textbf{Computer science}: 
source code,
automata theory,
computer security,
control flow,
boolean expressions,
double-precision floating-point format,
linear search,
software development,
hash functions,
cyberbullying.\\
Topics randomly sampled from \url{https://en.wikipedia.org/wiki/Glossary_of_environmental_science}.
\end{itemize}

\section{Methods for Study 2}

\subsection{Survey Questions}

\begin{enumerate}
    \item What year of your graduate program are you in?
    \item What kind of graduate program are you in?
    \item What discipline do you study?
    \item How often do you write about technical topics for a general audience? e.g. blog posts, opinion articles, essays, etc.
    \item How often do you post on Twitter about technical topics?
\end{enumerate}

\subsection{Interview Questions}

Questions about the task:

\begin{enumerate}
    \item Did you find any of the sparks helpful? If so, could you recall one spark that was helpful and explain in what way it helped?
(Make sure to dig into how the spark related to what they eventually wrote. Ask them to point it out in what they wrote.)
\item How do you think the sparks differed from what you would find on Wikipedia? How about Google search, or some other resource you use often?

\item How did the existing prompts differ from your custom prompts?

\item Could you recall one spark that wasn’t helpful, and explain why?

\item Were any of the sparks presented incorrect in some way? If so, what did you think of these?

\item What made you decide to stop generating sparks?

\item Did you have any concerns about ownership or agency?

\end{enumerate}

Debriefing questions:

\begin{enumerate}
    \item Is there anything you’d like to share that I didn’t ask about?

\item Is there anything you’d like to know or ask me?

\end{enumerate}

\end{document}